\documentclass[epj]{svjour}

\usepackage{graphics}
\usepackage{amssymb,amsmath}

\def\bbbz{{\Bbb Z}}

\def\diag{\mbox{diag}\,}
\def\rank{\mbox{rank}\,}
\def\tr{\mbox{tr}\,}
\def\re{{\rm Re}\,}
\def\im{{\rm Im}\,}

\def\fr#1{{\mathfrak{#1}}}
\def\openone{\leavevmode\hbox{\small1\kern-3.3pt\normalsize1}}

\begin{document}

\title{On $N$-wave Type Systems and Their Gauge Equivalent}

\author{V. S. Gerdjikov\inst{1}\thanks{e-mail: {\tt
gerjikov@inrne.bas.bg}},\, G.  G.  Grahovski\inst{1}\thanks{e-mail: {\tt
grah@inrne.bas.bg}} \and N. A. Kostov\inst{2} \thanks{e-mail: {\tt
nakostov@ie.bas.bg}}}

\institute{
Institute for Nuclear Research and Nuclear Energy,
72 Tzarigradsko chaussee, 1784 Sofia, Bulgaria \and
Institute of Electronics,
72 Tzarigradsko chaussee, 1784 Sofia, Bulgaria } %

\date{Received: October 20, 2001 / Revised version: date}

\abstract{ The class of nonlinear evolution equations (NLEE) -- gauge
equivalent to the $N$-wave equations related to the simple Lie algebra
$\fr{g} $ are derived and analyzed. They are written in terms of ${\cal
S}(x,t)\in \fr{g} $ satisfying $r=\rank \fr{g} $ nonlinear constraints. The
corresponding Lax pairs and the time evolution of the scattering data are
found. The Zakharov--Shabat dressing method is appropriately modified to
construct their soliton solutions.}  

\maketitle

\section{Introduction and Preliminaries} \label{intro}

It is well known \cite{ZaTa,FaTa} that  the Lax representation
(\ref{eq:1.3})
\begin{eqnarray}\label{eq:1.3}
[L(\lambda ), M(\lambda )]=0
\end{eqnarray}
is invariant under the group of gauge transformations.

One of the first nontrivial examples of gauge equivalent
nonlinear evolution equations (NLEE) is provided by the nonlinear
Schr\"odinger equation (NLSE) \cite{Laksh,ZaTa,FaTa}:
\begin{equation}\label{eq:nlse}
iu_t + u_{xx} + 2 |u^2|u(x,t) =0,
\end{equation}
and the Heisenberg ferromagnet equation (HFE):
\begin{eqnarray}\label{eq:hfe}
iS_{t}^{(0)} = {1  \over 2 } [  S^{(0)}(x,t), S_{xx}^{(0)}], \\
S^{(0)}(x,t) = g^{(0)}{}^{-1} \sigma _3 g^{(0)}(x,t); \nonumber
\end{eqnarray}
obviously $(S^{(0)})^2 =\openone  $. The equivalence between
(\ref{eq:nlse}) and (\ref{eq:hfe}) is based on the fact that
$g^{(0)}(x,t) $ is determined by $u(x,t) $ through (see \cite{FaTa}):
\begin{eqnarray}\label{eq:u-g0}
i {dg^{(0)}  \over dx } + q^{(0)}(x,t) g^{(0)}(x,t) =0, \\
q^{(0)}(x,t) = \left( \begin{array}{cc} 0 & u \\ -u^* & 0 \end{array}
\right), \qquad \lim_{x\to\infty }g^{(0)}(x,t)=\openone .
\end{eqnarray}

Both equations are infinite dimensional completely integrable Hamiltonian
systems. The phase space ${\cal  M}_{\rm NLSE} $ is the linear space of
all off-diagonal matrices $q^{(0)}(x,t) $ tending fast enough to zero for
$x\to\pm\infty $. A hierarchy of pair-wise compatible symplectic
structures on ${\cal  M}_{\rm NLSE} $ is provided by the $2 $-forms:
\begin{equation}\label{eq:ome-nls}
\Omega _{\rm NLSE}^{(k)} = {i \over 4} \int_{-\infty }^{\infty } dx \tr
\left( \delta q^{(0)} \wedge \Lambda ^k [\sigma _3, \delta q^{(0)}(x,t) ]
\right).
\end{equation}
The phase space ${\cal  M}_{\rm HFE} $ of the HFE is the manifold of all
$S^{(0)}(x,t) $ determined by the second relation in (\ref{eq:hfe}). The
family of compatible $2 $-forms is:
\begin{equation}\label{eq:ome-hf}
\tilde{\Omega} _{\rm HFE}^{(k)} = {i \over 4} \int_{-\infty }^{\infty }
dx \tr \left( \delta S^{(0)} \wedge \tilde{\Lambda} ^k [S^{(0)}, \delta
S^{(0)}(x,t) ] \right).
\end{equation}
By $\Lambda  $ and $\tilde{\Lambda } $ we have denoted the recursion
operator of the NLS type equations and its gauge equivalent \cite{vg-ya}.
The spectral theory of these two operators underlie all the fundamental
properties of these two classes of gauge equivalent NLEE, for details see
\cite{vg-ya}.  Note that the gauge transformation relates nontrivially the
symplectic structures, i.e. $\Omega _{\rm NLSE}^{(k)} \simeq\tilde{\Omega}
_{\rm HFE}^{(k+2)} $ \cite{KuRe,vg-ya}.

The NLSE is solvable by the inverse scattering method applied to the
Zakharov-Shabat system. It can be generalized to any simple Lie
algebra $\fr{g} $ of rank $r>1 $ by \cite{ZM,ZM1,K,1,ForKu*83,G*86}:
\begin{eqnarray}\label{eq:1.1}
L(\lambda )\psi  \equiv  \left(i {d  \over dx}+ q(x,t)-\lambda
J\right) \psi (x,t,\lambda )=0,
\end{eqnarray}
where $q(x,t) $ and $J $ are elements $\fr{g} $. A natural choice of the
gauge in (\ref{eq:1.1}) consists in choosing $J $ to be a
constant regular element of the Cartan subalgebra $\fr{h} $ of $\fr{g} $.
Next by a gauge transformation $L(\lambda )\to g_0^{-1}L(\lambda ) g_0 $
where $g_0(x,t) \in \fr{G} $ commutes with $J $ we can eliminate the
diagonal elements of $q(x,t) $ and cast it in the form $q(x,t)=[J,Q(x,t)]
$. Since $J $ is a regular element of $\fr{h} $ then $q(x,t) \in {\frak
g}\backslash {\frak  h}  $ is determined by $|\Delta | $ independent
coefficient functions as follows:
\begin{eqnarray}\label{eq:1.2}
Q(x,t)=\sum_{\alpha \in \Delta ^+} \left( q_{\alpha }(x,t) E_{\alpha } +
p_{\alpha }(x,t)E_{-\alpha }\right),
\end{eqnarray}
where $E_{\pm \alpha } $ are the root vectors of ${\frak  g} $ and
$\Delta _+ $ is the set of positive roots of ${\frak  g} $, $\Delta
=\Delta ^+ \cup (-\Delta ^+) $. The root $\alpha  $ is positive if $\alpha
(J)>0$; by $E_{\alpha }$,  $\alpha \in \Delta$ and $ H_i $, $i=1, \dots, r
$ we denote the Cartan--Weyl basis of ${\frak  g} $ with the standard
commutation relations, see \cite{Helg}.
This choice of the gauge of $L(\lambda ) $ allows one to study the
$N$-wave type equations:
\begin{eqnarray}\label{eq:1.5}
i[J,Q_t]-i[I,Q_x]+[[I,Q],[J,Q]]=0.
\end{eqnarray}

Indeed eq. (\ref{eq:1.5}) allows Lax representation with the pair  of Lax
operators $L(\lambda ) $ (\ref{eq:1.1}) and $M(\lambda )$:
\begin{eqnarray}\label{eq:1.4}
M(\lambda )\psi  \equiv \left( i {d  \over dt}+ [I,
Q(x,t)]-\lambda I\right) \psi (x,t,\lambda )=0,
\end{eqnarray}
where $I $ is also a constant regular element of $I\in {\frak  h}$.

There is a second canonical way to fix up the gauge of the Lax operator
known as the pole gauge \cite{Za*Mi}:
\begin{eqnarray}\label{eq:2.3}
\tilde{L}\tilde{\psi }(x,t,\lambda )\equiv \left(i{d  \over
dx}-\lambda {\cal  S}(x,t) \right) \tilde{\psi }(x,t,\lambda )=0,
\end{eqnarray}
where $\tilde{\psi }(x,t,\lambda ) = g^{-1}(x,t)\psi (x,t,\lambda )$,
\begin{eqnarray}\label{eq:2.4}
{\cal  S}(x,t) = \mbox{Ad}_{g}\cdot J \equiv g^{-1}(x,t)Jg(x,t).
\end{eqnarray}
The gauge transformation which takes $L(\lambda ) $ to $\tilde{L}(\lambda
) =g^{-1}L(\lambda ) g(x,t) $ is performed with the Jost solution of
(\ref{eq:1.1}) taken at $\lambda =0 $, i.e. $g(x,t)\in \fr{G} $ and
\begin{equation}\label{eq:gxt}
i {dg  \over dx } +q(x,t) g(x,t) =0, \qquad \lim_{x\to\infty } g(x,t)
=\openone .
\end{equation}
Applying the gauge transformation to $M(\lambda ) $ we get:
\begin{equation}\label{eq:}
\tilde{M}\tilde{\psi }(x,t,\lambda )\equiv \left(i{d  \over dt}-\lambda
f({\cal  S}) \right) \tilde{\psi }(x,t,\lambda)=0,
\end{equation}
where $f({\cal  S}) $ is a function (in fact, a polynomial) to be
determined below. Indeed only this choice of $\tilde{M}(\lambda ) $
ensures the vanishing of the $\lambda ^2 $-term in the zero-curvature
condition $[\tilde{L}, \tilde{M}]=0 $; the terms proportional to $\lambda
$ lead to the following form of the gauge equivalent NLEE:
\begin{eqnarray}\label{eq:2.5}
{\cal  S}_t- {d  \over dx }f({\cal  S})=0,
\end{eqnarray}

While the $N $-wave type equations are well known their gauge equivalent
ones to the best of our knowledge have not been derived yet. One of the
difficulties in doing this is the necessity to express all factors in
terms of ${\cal  S} $ only.

The gauge equivalent operators $L(\lambda ) $ and $\tilde{L}(\lambda ) $
have equivalent spectral properties and spectral data. This fact allows
one to prove that the classes of NLEE related to $L(\lambda ) $ and
$\tilde{L}(\lambda ) $ are also equivalent.

In Section 2 we construct the NLEE gauge equivalent to the
$N$-wave equations (\ref{eq:1.5})  extending the results in
\cite{vg-ya}.  Namely we calculate the functions $f({\cal  S}) $
for the cases when $\fr{g} $ belongs to the classical series of
simple Lie algebras. In Section 3 we briefly describe the
interrelations between the scattering data of $L(\lambda ) $ and
$\tilde{L}(\lambda ) $ and outline some of their reductions. We
also reformulate the Riemann--Hilbert problem (RHP) for the gauge
equivalent systems and describe the time evolution of the
scattering data. In section 4 we extend the Zakharov--Shabat
dressing method \cite{ZM1,1,Za*Mi} for the gauge equivalent
systems \cite{vg-ya} and provide the general form of the 1-soliton
solution of these system.  These results are demonstrated on an
example on the orthogonal Lie algebra ${\bf B}_2\simeq so(5) $ in
section 5. This and other particular cases of Eqs. (\ref{eq:2.5})
describe isoparametric hypersurfaces \cite{Ferap}.

\section{General form of the gauge equivalent systems}
\label{sec:1}

It is natural that $f({\cal S})=g^{-1}(x,t) I g(x,t) $, i.e., it is
uniquely determined by $I $. Both $J $ and $I $ belong to the Cartan
subalgebra $\fr{h} $ so they have common set of eigenspaces.

The derivation of the corresponding functions $f({\cal  S}) $ is different
for ${\bf A}_r $ and for ${\bf B}_r $, ${\bf C}_r $, ${\bf D}_r $ series.
Let first $\fr{g}\simeq {\bf A}_r = sl(n) $ with $n=r+1 $. Then
\[
J= \diag (J_1,\dots, J_n) , \qquad I= \diag (I_1,\dots, I_n),
\]
and the only constraint on the eigenvalues $J_k $ and $I_k $ is $\tr J =
\tr I =0 $. The projectors on the common eigensubspaces of $J $ and $I $
are given by:
\begin{equation}\label{eq:pi_k}
\pi_k (J) = \prod_{s\neq k} {J-J_s  \over J_k-J_s } =
\diag (0,\dots, 0,\mathop{1}\limits_k , 0,\dots, 0).
\end{equation}
Next we note that
\begin{equation}\label{eq:I-sln}
I= \sum_{k=1}^{n} I_k \pi_k(J).
\end{equation}
In order to derive $f({\cal  S}) $ for $\fr{g}\simeq sl(n) $ we need to
apply the gauge transformation to (\ref{eq:I-sln}) with the result:
\begin{equation}\label{eq:f-sln}
f({\cal  S}) = \sum_{k=1}^{n} I_k \pi_k({\cal  S}),
\end{equation}
i.e., $f({\cal  S})  $ is a polynomial of order $n-1 $. Obviously ${\cal
S} $ is restricted by:
\begin{equation}\label{eq:S-cha}
\prod_{k=1}^{n} ({\cal  S} - J_k) =0, \qquad \tr {\cal  S}^k = \tr J^k,
\end{equation}
for $k=2,\dots,n $.

Let us now assume that $\fr{g} $ is an orthogonal or symplectic algebra.
In the typical representation we can introduce the Cartan generators
$H_{e_k} $ which are dual to the orthogonal basic vectors $e_k $ in the
root space. Each $H_{e_k} $ has only two non-vanishing eigenvalues equal
to $1 $ and $-1 $ respectively. Then we put:
\[ J=\sum_{k=1}^{r}J_k H_{e_k} , \qquad I=\sum_{k=1}^{r}I_k H_{e_k} .
\]
Obviously the odd powers of $H_{e_k} $ also belong to $\fr{g} $ while the
even powers do not. The projectors $f_k(J) $ onto $H_{e_k} $ then can be
written down as:
\begin{equation}\label{eq:f_k}
f_k(J) = {J  \over J_k } \prod_{s\neq k}^{} {J^2 -J_s^2 \over
J_k^2 - J_s^2 } = H_{e_k} \in \fr{h}.
\end{equation}
Therefore
\begin{eqnarray}\label{eq:I}
I&=& \sum_{k=1}^{r} I_k f_k(J),
\end{eqnarray}
and applying the gauge transformation we get:
\begin{eqnarray}\label{eq:f(S)}
f({\cal  S})&\equiv & g^{-1}(x,t) I g(x,t) = \sum_{k=1}^{r} I_k
f_k({\cal  S}).
\end{eqnarray}

Then the equation gauge equivalent to (\ref{eq:1.1}) is given by
(\ref{eq:2.5}) with $f({\cal  S}) $ determined by (\ref{eq:f(S)}).

In addition ${\cal  S}(x,t) $ satisfies a set of nonlinear constraints;
one of them is the characteristic equation:
\begin{eqnarray}\label{eq:2.6}
{\cal  S}^{\kappa _0}\prod_{k=1}^{r}({\cal  S}^2-J_k^2)=0,
\end{eqnarray}
where $\kappa _0= 0$ if $ {\frak  g} \simeq C_r$ or $D_r$ and $\kappa _0=1
$, if ${\frak  g} \simeq B_r$.
To construct the others we use the typical representation of $\fr{g} $.
It this settings we see that all even powers of $H_{e_k} $
have trace equal to $2 $. Thus we have:
\begin{equation}\label{eq:tr_Jk}
\tr (J^{2k}) \equiv 2 \sum_{p=1}^{r} J_p^{2k} = \tr ({\cal  S})^{2k},
\end{equation}
for $k=1,\dots,r $. The conditions (\ref{eq:tr_Jk}) are precisely
$r $ independent algebraic constraints on ${\cal  S} $. Solving
for them we conclude that the number of independent coefficients
in ${\cal  S} $ is equal to the number of roots $|\Delta | $ of
$\fr{g} $.

Both classes of NLEE possess hierarchies of Hamiltonian structures. The
phase space ${\cal  M}_{\rm N-w} $ of the $N $-wave equations is the
linear space of off-diagonal matrices $q(x,t) $; the hierarchy of
symplectic structures is given by:
\begin{equation}\label{eq:ome-nw}
\Omega _{\rm N-w}^{(k)} = i \int_{-\infty }^{\infty } dx \tr
\left( \delta q \wedge \Lambda ^k [J, \delta q(x,t) ]\right).
\end{equation}
The phase space ${\cal  M}_{\cal S } $ of their gauge equivalent equations
(\ref{eq:2.5}) is the nonlinear  manifold of all ${\cal  S}(x,t) $
satisfying equations (\ref{eq:2.6}), (\ref{eq:tr_Jk}). The family of
compatible $2 $-forms is:
\begin{equation}\label{eq:ome-S}
\tilde{\Omega} _{\cal S}^{(k)} = i \int_{-\infty }^{\infty } dx \tr \left(
\delta {\cal  S} \wedge \tilde{\Lambda} ^k [{\cal  S}, \delta {\cal
S}(x,t) ] \right).
\end{equation}
Here $\Lambda  $ and $\tilde{\Lambda } $ are the recursion operator of the
$N $-wave type equations (see \cite{vg-ya}) and its gauge equivalent:
$\tilde{\Lambda }=g^{-1}\Lambda g(x,t) $.

\section{Fundamental analytic solutions (FAS) and scattering data for gauge
equivalent systems}
\label{sec:2}

The direct scattering problem for the Lax operator (\ref{eq:1.1}) is based
on the Jost solutions:
\begin{eqnarray}\label{eq:1.6}
\lim_{x \to \infty }\psi (x,\lambda )e^{i\lambda Jx}=\openone,\qquad
\lim_{x \to -\infty }\phi (x,\lambda )e^{i\lambda Jx}=\openone,
\end{eqnarray}
and the scattering matrix:
\begin{eqnarray}\label{eq:1.7}
T(\lambda )= (\psi (x,\lambda ))^{-1}\phi (x,\lambda ).
\end{eqnarray}
The FAS $\xi ^{\pm}(x,\lambda ) $ of $
L(\lambda ) $ are analytic functions of $\lambda  $ for $\lambda
\gtrless 0 $ and are related to the Jost solutions by \cite{G*86}
\begin{eqnarray}\label{eq:1.8}
\xi^{\pm}(x,\lambda ) = \phi(x,\lambda )S^{\pm}(\lambda ) =
\psi (x,\lambda )T^{\mp}(\lambda )D^{\pm }(\lambda ),
\end{eqnarray}
where $T^{\pm}(\lambda ) $, $S^{\pm}(\lambda ) $ and
$D^{\pm}(\lambda ) $ are the factors of the Gauss decomposition of the
scattering matrix:
\begin{eqnarray}\label{eq:1.9}
T(\lambda )= T^-(\lambda )D^+(\lambda )\hat{S}^+(\lambda )=
T^+(\lambda )D^-(\lambda )\hat{S}^-(\lambda ).
\end{eqnarray}
Here $\hat{S}\equiv S^{-1} $, the superscripts $+$ (resp. $-$) in $T^{\pm}
(\lambda ) $ and $S^{\pm}(\lambda ) $ mean upper- (resp. lower-)
triangularity. The diagonal factors $ D^{\pm}(\lambda ) $  are analytic
functions of $\lambda $ for $\mbox{Im}\, \lambda >0 $ and $\mbox{Im}\,
\lambda <0 $ respectively.

On the real axis $\xi ^+(x, \lambda ) $ and $\xi ^-(x, \lambda ) $ are
related by
\begin{equation}\label{eq:1.10}
\xi ^+(x,\lambda ) =\xi ^-(x,\lambda ) G_0(\lambda ), \quad
G_0(\lambda )= \hat{S}^-(\lambda ) S^+(\lambda ),
\end{equation}
and the function $G_0(\lambda ) $ can be considered as a minimal set of
scattering data in the case of absence of discrete eigenvalues of
(\ref{eq:1.1}) \cite{Sh,G*86}.

If the potential $q(x,t) $ of $L(\lambda ) $ (\ref{eq:1.1}) satisfies
equation (\ref{eq:1.5}) then $S^{\pm}(\lambda ) $ and $T^{\pm}(\lambda )
$ satisfy the linear equation:
\begin{eqnarray}\label{eq:1.11}
i{dS^{\pm} \over dt}-\lambda [I,S^{\pm}]=0 \qquad
i{dT^{\pm} \over dt}-\lambda [I,T^{\pm}]=0,
\end{eqnarray}
while the functions $D^{\pm}(\lambda ) $ are time-independent. In other
words $D^{\pm}(\lambda ) $ can be considered as the generating functions
of the integrals of motion of (\ref{eq:1.5}).

In order to determine the scattering data for the gauge equivalent
equations we need the FAS for these systems:
\begin{eqnarray}\label{eq:3.1}
\tilde{\xi }^{\pm}(x,\lambda )=g^{-1}(x,t)\xi ^{\pm}(x,\lambda )g_-,
\end{eqnarray}
where $g_-= \lim_{x \to -\infty }g (x,t) =\hat{T}(0)$.  In order to
ensure that the
functions $\tilde{\xi }^{\pm}(x,\lambda ) $ are analytic with respect to
$\lambda $ the scattering matrix $T(0) $ at $ \lambda =0 $ must belong to
the corresponding Cartan subgroup ${\frak  H} $. Then Equation
(\ref{eq:3.1}) provide the FAS of $\tilde{L} $.
We calculate their asymptotics for $x\to\pm\infty  $ and
establish the relations between the scattering matrices of the two
systems:
\begin{eqnarray}\label{eq:3.2}
&&\lim_{x \to -\infty }\tilde{\xi }^+(x,\lambda )=
T(0)S^+(\lambda )\hat{T}(0) \\
&&\lim_{x \to \infty }\tilde{\xi }^+(x,\lambda )=
e^{-i\lambda Jx}T^-(\lambda )D^+(\lambda )\hat{T}(0)
\end{eqnarray}
with the result:
\begin{eqnarray}\label{eq:3.3}
\tilde{T}(\lambda )= T(\lambda )\hat{T}(0).
\end{eqnarray}
Obviously  $\tilde{T}(0)=\openone  $.
The factors in the corresponding Gauss decompositions are related by:
\begin{eqnarray}\label{eq:3.4}
\tilde{S}^{\pm}(\lambda )= T(0)S^{\pm}(\lambda )\hat{T}(0), \qquad
\tilde{T}^{\pm}(\lambda )=T^{\pm}(\lambda ) \nonumber\\
\tilde{D}^{\pm}(\lambda )=D^{\pm}(\lambda )\hat{T}(0).
\end{eqnarray}

On the real axis $\tilde{\xi }^+(x,\lambda ) $ and
$\tilde{\xi }^-(x,\lambda ) $ are related by:
\begin{eqnarray}\label{eq:3.5}
\tilde{\xi }^+(x,\lambda )=\tilde{\xi }^-(x,\lambda )\tilde{G}_0(\lambda
), \qquad \tilde{G}_0(\lambda )=\hat{\tilde{S}}^-(\lambda )
\tilde{S}^+(\lambda )
\end{eqnarray}
with the normalization condition $\tilde{\xi }(x,0)=\openone $;
again $\tilde{G}_0(\lambda ) $ can be considered as a minimal set of
scattering data.

The numerous $\bbbz_2 $-reductions that have been recently classified for
the $N $-wave equations \cite{vgrn,vgn} using the reduction group
introduced by Mikhailov \cite{2}. They can easily
be reformulated for the gauge equivalent systems. Here we write down only
two of them:
\begin{eqnarray}\label{eq:rd}
&& \mbox{1)} \qquad {\cal  S}^\dag (x,t) = K {\cal  S}(x,t) K^{-1}, \qquad
K\in \fr{H}, \quad K^2=\openone , \nonumber\\
&& \mbox{2)} \qquad {\cal  S}(x,t) =  {\cal
S}(x,t)^T.
\end{eqnarray}
Obviously each of the constraints 1) and 2) are
compatible with equation (\ref{eq:2.5}) and diminishes the number of
independent coefficients by a factor of 2.

\section{Dressing factors and 1-soliton solutions}
\label{sec:3}

The main idea of the dressing method is starting from a FAS $\tilde{\xi
}_{(0)}^{\pm}(x,\lambda ) $ of $\tilde{L} $ with potential ${\cal
S}_{(0)} $ to construct a new singular solution   $\tilde{\xi
}_{(1)}^{\pm}(x,\lambda ) $ of the RHP (\ref{eq:3.5}) with singularities
located at prescribed positions $\lambda _1^{\pm} $. Then the new
solutions $\tilde{\xi }_{(1)}^{\pm}(x,\lambda ) $ will correspond to a
potential ${\cal  S}_{(1)} $ of $\tilde{L} $ with two discrete eigenvalues
$\lambda _1^{\pm} $. It is related to the regular one by the dressing
factors $\tilde{u}(x,\lambda ) $:
\begin{eqnarray}\label{eq:4.1}
\tilde{\xi }_{(1)}^{\pm}(x,\lambda ) &=& \tilde{u}(x,\lambda )
\tilde{\xi }_{(0)}^{\pm}(x,\lambda ) \tilde{u}_-^{-1}(\lambda),\nonumber\\
\tilde{u}_-(\lambda )&=&\lim_{x \to -\infty }\tilde{u}(x,\lambda ),
\end{eqnarray}
and the dressing factors for the gauge equivalent equations
$\tilde{u}(x,\lambda ) $ are related to $u(x,\lambda ) $ by
\begin{eqnarray}\label{eq:4.2}
\tilde{u}(x,\lambda ) = g_{(0)}^{-1}(x,t)u^{-1}(x,\lambda =0) u(x,\lambda
)g_{(0)}.
\end{eqnarray}
If ${\frak  g}\simeq {\bf A}_r $ then the gauge equivalent
dressing factors are
\begin{eqnarray}\label{eq:4.3}
\tilde{u}(x,\lambda )&=& \openone +\left({c_1(\lambda )  \over c_1(0)
}-1\right) P_1, \qquad c_1(\lambda )={\lambda -\lambda _1^+  \over \lambda
-\lambda _1^- } \nonumber\\
P_1(x) &= &{|n(x)\rangle \langle m(x)  \over \langle m(x)|n(x)\rangle  },
\\
|n(x)\rangle &=& \xi _0^+(\lambda _1^+)|n_0\rangle , \qquad \langle m(x)|=
\langle m_0|\hat{\xi} _0^-(\lambda _1^-), \nonumber
\end{eqnarray}
where $|n_0\rangle  $ and $\langle m_0| $ are constant vectors and these
dressing factors satisfy the equation:
\begin{eqnarray}\label{eq:4.5}
i {d\tilde{u} \over dx }-\lambda {\cal  S}_{(1)}\tilde{u}+\lambda
\tilde{u}{\cal S}_{(0)}=0.
\end{eqnarray}
If ${\frak  g}\simeq {\bf B}_r, {\bf D}_r $ the dressing factors
take the form \cite{vgrn}:
\begin{eqnarray}\label{eq:4.6}
&&u(x,\lambda )=\openone + (c_1(\lambda )-1)P_1+
(c_1^{-1}(\lambda )-1)P_{-1} \\
&&\tilde{u}(x,\lambda )=\openone +\left({c_1(\lambda )  \over c_1(0) }
-1\right)\tilde{P}_1 +\left({c_1(0)  \over c_1(\lambda ) } -1\right)
\tilde{P}_{-1},
\end{eqnarray}
where $P_{-1}(x)=SP_1^T(x)S^{-1} $, $P_1(x) $ is the rank 1 projector
(\ref{eq:4.3}), $\tilde{P}_{\pm 1}=g_{(0)}^{-1}P_{\pm 1}g_{(0)}(x,t) $.
If ${\frak  g}\simeq B_r$ then $N=2r+1 $,
\begin{eqnarray}\label{eq:4.7}
S= \sum_{k=1}^{r}(-1)^{k+1}(E_{k\bar{k}}+ E_{\bar{k}k}) +
(-1)^rE_{r+1,r+1};
\end{eqnarray}
$\bar{k}=N-k+1$, $(E_{km})_{il}=\delta _{ik}\delta _{ml} $;
if ${\frak  g}\simeq D_r$ then $  N=2r $ 
\begin{eqnarray}\label{eq:4.7'}
S= \sum_{k=1}^{r}(-1)^{k+1}(E_{k\bar{k}}+ E_{\bar{k}k}),
\end{eqnarray}
 If the dressing factors
of the gauge equivalent equations satisfy (\ref{eq:4.5}) then the
projectors $\tilde{P}_{\pm 1} $ satisfy the equations:
\begin{eqnarray}\label{eq:4.8}
i{d\tilde{P}_1  \over dx }+ \lambda _1^-\tilde{P}_1{\cal S}_{(0)}-\lambda
_1^-{\cal S}_{(1)}\tilde{P}_1=0, \nonumber\\
i{d\tilde{P}_{-1}  \over dx }+ \lambda _1^+\tilde{P}_{-1}{\cal S}_{(0)}-
\lambda _1^+{\cal S}_{(1)}\tilde{P}_{-1}=0,
\end{eqnarray}
and the "dressed" potential can be obtained by:
\begin{eqnarray}\label{eq:4.9}
{\cal  S}_{(1)}=S_{(0)}+i {\lambda _1^+- \lambda _1^- \over \lambda _1
^+ \lambda _1^- } {d  \over dx }(\tilde{P}_1(x)-\tilde{P}_{-1}(x)).
\end{eqnarray}
The dressing factors can be written in the form:
\begin{eqnarray}\label{eq:4.10}
\tilde{u}(x,\lambda )=\exp \left[ \ln \left({c_1(\lambda )  \over c_1(0) }
\right)\tilde{p}(x)\right],
\end{eqnarray}
where $ \tilde{p}(x)=\tilde{P}_{1}-\tilde{P}_{-1} \in {\frak  g} $ and
consequently $\tilde{u}(x,\lambda ) $ belongs to the corresponding
orthogonal group.

Making use of the explicit form of the projectors $P_{\pm 1}(x) $ valid
for the typical representation of ${\bf B}_r $ we have \cite{vgrn}
\begin{eqnarray}\label{eq:4.11}
\tilde{p}(x)&=&{2  \over \langle m|n\rangle  }\sum_{k=1}^{r}
\tilde{h}_k(x)H_{e_k} \nonumber\\
&+&{2  \over \langle m|n\rangle }\sum_{\alpha \in \Delta _+}
(\tilde{P}_{\alpha }(x)E_{\alpha }+\tilde{P}_{-\alpha }(x)E_{-\alpha }),
\end{eqnarray}
where we assumed ${\cal  S}_{(0)}=J $, $g_{(0)}=\openone  $. Thus
\begin{eqnarray}\label{eq:4.13a}
\tilde{h}_k(x,t)&=&n_{0,k}m_{0,k}e^{2\nu _1y_k}- n_{0,\bar{k}}
m_{0,\bar{k}}e^{-2\nu _1y_k}, \nonumber\\
\langle m|n\rangle &=& \sum_{k=1}^{r}(n_{0,k}m_{0,k}e^{2\nu _1y_k}
+n_{0,\bar{k}}m_{0,\bar{k}}e^{-2\nu _1y_k}) \nonumber\\
&+&n_{0,r+1}m_{0,r+1}, \\
\label{eq:4.13b}
\tilde{P}_\alpha &=&\left\{ \begin{array}{ll}
\tilde{P}_{ks}, & \qquad \mbox{for }\; \alpha =e_k-e_s \\
\tilde{P}_{k\bar{s}}, & \qquad \mbox{for }\; \alpha =e_k+e_s \\
\tilde{P}_{k,r+1}, & \qquad \mbox{for }\; \alpha =e_k \\
\end{array} \right. \nonumber
\end{eqnarray}
Here $1\leq k , s \leq r $, $\mu _1=\re \lambda _1^+ $, $\nu _1=\im
\lambda _1^+ $ and
\begin{eqnarray}\label{eq:4.14}
\tilde{P}_{ks} = e^{i\mu _1(y_s-y_k)}(n_{0,k}m_{0,s}e^{\nu _1(y_s+y_k)}
\nonumber\\
- (-1)^{k+s}n_{0,\bar{s}}m_{0,\bar{k}}e^{-\nu _1(y_s+y_k)}), \nonumber\\
y_k=J_kx+I_kt, \qquad y_{\bar{k}}=-y_k, \qquad y_{r+1}=0.
\end{eqnarray}
The corresponding result for the ${\bf D}_r $ series is obtained
formally if in the above expressions (\ref{eq:4.13a}),
(\ref{eq:4.13b}) and (\ref{eq:4.14}) we put
$n_{0,r+1}=m_{0.r+1}=0 $. Thus
$\tilde{P}_{k,r+1}=\tilde{P}_{r+1,k}=0 $ and the last term in the
right hand side of $\langle m|n\rangle  $ (\ref{eq:4.13a}) is
missing.

The reductions (\ref{eq:rd}) applied to the
$1 $-soliton solution constraint the vectors $|n_0\rangle  $, $\langle
m_0| $ and the eigenvalues $\lambda _1^\pm $:
\begin{eqnarray}\label{eq:rdn}
&& \mbox{1)} \qquad |n_0\rangle = K |m_0^*\rangle ,\qquad \lambda _1^- =
(\lambda _1^+)^*, \nonumber\\
&& \mbox{2)} \qquad |n_0\rangle =  |m_0\rangle , \qquad \lambda _1^- =
-\lambda _1^+.
\end{eqnarray}
The $N $-soliton solutions can be obtained by applying successively $N $
times the dressing procedure.

\section{Example: ${\frak  g}\simeq {\bf B}_2 $ algebra}
\label{sec:4}

This algebra has four positive roots: $e_1\pm e_2$, $e_1 $ and $e_2 $.
The corresponding $4 $-wave system has the form:
\begin{eqnarray}\label{eq:b2.4}
&& i(J_1-J_2)q_{10,t}-i(I_1-I_2)q_{10,x}+2\kappa q_{11}q_{01}^*=0,
\nonumber\\
&&iJ_2q_{01,t}-iI_2q_{01,x}+\kappa (q_{11}^*q_{12}+q_{11}q_{10}^*)=0,
\nonumber\\
&&iJ_1q_{11,t}-iI_1q_{11,x}+\kappa (q_{12}q_{01}^*-q_{10}q_{01})=0, \\
&&i(J_1+J_2)q_{12,t}-i(I_1+I_2)q_{12,x}-2\kappa q_{11}q_{01}=0. \nonumber
\end{eqnarray}
where $\kappa =J_1I_2-J_2I_1 $ and the subscripts $10 $, $01 $, $11 $ and
$12 $ refer to the roots $e_1-e_2 $, $e_1 $, $e_2 $ and $e_1+e_2 $
respectively.
This system has applications in nonlinear optics \cite{1,vgrn} and in
differential geometry \cite{Ferap}. Its gauge equivalent is:
\begin{eqnarray}\label{eq:ex1}
&& {\cal  S}_t- f_1 {\cal  S}_x - f_3 ({\cal  S}^3)_x=0, \nonumber\\
&& f_1= {I_2J_1^3-I_1J_2^3  \over J_1J_2(J_1^2-J_2^2) } \qquad
f_3= {I_1J_2-I_2J_1  \over J_1J_2(J_1^2-J_2^2) },
\end{eqnarray}
where the $5\times 5 $ matrix ${\cal  S} $ is constrained by:
\begin{eqnarray}\label{eq:S-cons}
\tr {\cal  S}^2 = 2(J_1^2 + J_2^2), \qquad
\tr {\cal  S}^4 = 2(J_1^4 + J_2^4), \nonumber\\
{\cal  S} ({\cal  S}^2 - J_1^2) ({\cal  S}^2 - J_2^2) =0.
\end{eqnarray}

We write down the $1 $-soliton solution for a special choice
\begin{eqnarray}\label{eq:1s}
&& n_{0,1}=1, \qquad n_{0,2}=\rho , \qquad n_{0,3} = \sqrt{2(\rho ^2-1)},
\nonumber\\
&& n_{0,k}= n_{0,\bar{k}}, \qquad m_{0,k}=n_{0,k}.
\end{eqnarray}
of the soliton parameters with $\rho \geq 1 $ and real.
The choice (\ref{eq:1s}) satisfies (\ref{eq:rdn}) with $K=\openone  $.
Inserting it into the general formulae
(\ref{eq:4.11})--(\ref{eq:4.13b}) we get $\tilde{P}_{-\alpha } =
\tilde{P}_{\alpha } ^* $ with:
\begin{eqnarray}\label{eq:1ss}
\langle m|n\rangle &=& 2(\sinh^2 \nu _1 y_1 + \rho ^2 \cosh^2 \nu _1 y_2),
\nonumber\\
\tilde{h}_1 &=& \sinh 2\nu _1 y_1, \qquad
\tilde{h}_2 = \rho ^2 \sinh 2\nu _1 y_2, \nonumber\\
\tilde{P}_{e_1 \pm e_2} &=& \rho e^{-i\mu _1(y_1 \pm y_2)} \cosh \nu _1
(y_1 \mp y_2), \\
\tilde{P}_{e_k} &=& \sqrt{2 (\rho^2-1)} e^{-i\mu _1y_k} \sinh \nu _1 y_k ,
\qquad k=1,2. \nonumber
\end{eqnarray}
If $\rho =1 $ we get a $1 $-soliton solution associated with the $
D_2 \simeq A_1\oplus A_1 $ subalgebra; $\rho =0 $ gives
a $1 $-soliton solution for the $so(3) $ subalgebra of $B_2 $.
In both subcases the subsets of roots ($\{\pm e_1 \pm e_2\} $ and $\{ \pm
e_1, \pm e_2\} $ resp.) for which $\tilde{P}_{\alpha } \not=0 $ contain
only roots with the same length.

\section{Discussion}
\label{sec:5}

We derived the explicit form of the NLEE gauge equivalent to the $N $-wave
equations related to the classical simple Lie algebras $\fr{g} $. These
equations are Hamiltonian ones. Their phase space $\tilde{{\cal  M}} $ is
a nonlinear one, since ${\cal  S} $ is restricted by the nonlinear
constraints (\ref{eq:2.6}), (\ref{eq:tr_Jk}).

The gauge covariant formulation of the spectral decompositions of the
recursion operators $\Lambda  $ and its gauge equivalent $\tilde{\Lambda }
$ allows one to expect that the hierarchies of symplectic structures will
satisfy in analogy with the NLSE-HFE case the relation
$\Omega _{\rm N-w}^{(k)} \simeq\tilde{\Omega} _{\cal S}^{(k+2)} $.

Another open problem is the study of the
$\bbbz_2 $-reductions of (\ref{eq:2.5}) along the ideas outlined in
\cite{vgrn,vgn}.

\section*{Acknowledgements}
\label{sec:ack}

The authors are grateful to Professors V. Sokolov and E. Ferapontov for
useful discussions.

\end{document}